\documentclass[twocolumn]{aastex631}

\usepackage{amsmath}

\begin{document}

\title[Polarization of Quasar 1604+159]{Magnetic Field of the Quasar 1604+159 from Parsec to Kilo-parsec Scale }

\author[0000-0002-5398-1303]{Xu-Zhi Hu}
\thanks{E-mail: hxz@shao.ac.cn}
\affiliation{Shanghai Astronomical Observatory, Chinese Academy of Sciences, Shanghai 200030, People's Republic of China}
\affiliation{School of Physical Science and Technology, ShanghaiTech University, Shanghai 201210, People's Republic of China}
\affiliation{School of Astronomy and Space Science, University of Chinese Academy of Sciences, Beijing 100049, People's Republic of China}

\author[0000-0002-1992-5260]{Xiaoyu Hong}
\affiliation{Shanghai Astronomical Observatory, Chinese Academy of Sciences, Shanghai 200030, People's Republic of China}
\affiliation{School of Physical Science and Technology, ShanghaiTech University, Shanghai 201210, People's Republic of China}
\affiliation{School of Astronomy and Space Science, University of Chinese Academy of Sciences, Beijing 100049, People's Republic of China}
\affiliation{Key Laboratory of Radio Astronomy, Chinese Academy of Sciences, 210008 Nanjing, People's Republic of China}

\author[0000-0002-1992-5260]{Wei Zhao}
\affiliation{Shanghai Astronomical Observatory, Chinese Academy of Sciences, Shanghai 200030, People's Republic of China}
\affiliation{Key Laboratory of Radio Astronomy, Chinese Academy of Sciences, 210008 Nanjing, People's Republic of China}

\author[0000-0002-1908-0536]{Liang Chen}
\affiliation{Shanghai Astronomical Observatory, Chinese Academy of Sciences, Shanghai 200030, People's Republic of China}
\affiliation{Key Laboratory for Research in Galaxies and Cosmology, Shanghai Astronomical Observatory, Chinese Academy of Sciences, Shanghai 200030,\\
People’s Republic of China}

\author[0000-0001-9036-8543]{Wei-Yang Wang}
\affiliation{School of Astronomy and Space Science, University of Chinese Academy
of Sciences, Beijing 100049, People's Republic of China}

\author[0000-0003-3454-6522]{Linhui Wu}
\affiliation{Shanghai Astronomical Observatory, Chinese Academy of Sciences, Shanghai 200030, People's Republic of China}



\begin{abstract}
We present a multi-frequency polarimetric study for the quasar 1604+159.
The source was observed at the $L$ band with the American Very Long Baseline Array (VLBA) and the $L$, $X$, and $U$ bands with the Very Large Array (VLA).
These observations provide different resolutions from mas to arcsec, enabling us to probe the morphology and magnetic field from tens of parsec to hundreds of kilo-parsec scale.
We detect a symmetrical Fanaroff-Riley-Class-I-like structure.
The source has several lobes and bulges, forming a cocoon shape.
The polarization is normal to the edges of the structure with high fractional polarization up to $\sim 60\%$.
Two hotspots are observed at the eastern and western sides of the source, located symmetrically relative to the core.
The flux density ratio ($>1.5$) between the two hotspots suggests the Doppler beaming effect exists at a large scale.
The polarized emission in the hotspots also shows a symmetrical structure with an oblique direction from the jet direction.
In general, the jet propagates in a collimating structure with several bends.
Polarization is also detected perpendicular to the local jet from $\sim$100 mas to $\sim$ 1 arcsec.
The jet shows strong polarized intensity and high fractional polarization at the bending edges.
We discuss the possible origins of the observed structure and magnetic field.

\end{abstract}



\section{Introduction} \label{sec:intro}
Jets in active galactic nuclei (AGNs) result from the accretion onto the supermassive black holes (SMBHs).
For some active galaxies, jets carrying enormous energy serve as a direct physical link to the co-evolution between SMBHs and their hosting galaxies, in which low-mass SMBHs are in small galaxies and massive SMBHs are in massive galaxies. \citep{1998AJ....115.2285M,2014ARA&A..52..589H}.
It is believed that the propagation of jets as the energy released by supermassive black holes restrains the growth of massive galaxies, called jet-driven AGN feedback \citep{2012rjag.book.....B}.
Thus, the study of the propagation of jets contributes to the jet-driven feedback and the co-evolution, which is a multiscale physics.

Magnetic (\textit{\textbf{B}}) fields wound up by rotating accretion disks or central SMBHs extend with jets to a distance of parsec (pc) scale even up to kilo-parsec (kpc) scale \citep{2015A&A...583A..96G}.
\textit{\textbf{B}}-fields and ambient mediums regulate the jet structure (including the knots, hotspots, and lobes) and jet power.
Therefore, it is crucial to probe how magnetic fields and jets evolve with distance after the jet launching. 

Magnetic fields can be detected by linear polarization observation \citep[e.g.][]{1999ApJ...518L..87A,2000MNRAS.319.1109G,2010ApJ...723.1150P,2014MNRAS.438L...1G,2023MNRAS.519.2773B,2024MNRAS.527..672B}.
As the main radio emission mechanism of jets of AGN, synchrotron emission has a theoretical maximum fractional linear polarization of $75\%$, depending on the electron index \citep{1970ranp.book.....P}.
The projected components of $\textit{\textbf{B}}$-field onto the plane of the sky are orthogonal to the polarization angle $\chi$ in optically thin regions, while parallel in optically thick regions \citep{1970ranp.book.....P}.

The detection of Rotation Measure (RM) also provides information on the $\textit{\textbf{B}}$-fields \citep[]{2002PASJ...54L..39A,2005ApJ...626L..73Z,2012AJ....144..105H,2020ApJ...903...36S}.
Further, transverse RM gradients across jets provide strong evidence of helical $\textit{\textbf{B}}$-field \citep[e.g.][]{2004MNRAS.351L..89G,2008ApJ...682..798A,2010ApJ...720...41A,2008MNRAS.384.1003G,2009ApJ...694.1485K,2009MNRAS.400....2M,2010MNRAS.402..259C, 2012AJ....144..105H,2015MNRAS.450.2441G,2018A&A...612A..67G,2015A&A...583A..96G,2017Galax...5...61K}.

For objects with their jet axes orienting close to the line of sight (LOS) of the observer, emission from jets experience an enhancement of power due to the Doppler boosting effect, which makes them natural candidates to study $\textit{\textbf{B}}$-fields in jets of AGNs, thus the jet-driven feedback.
This requires comprehensive studies of magnetic fields and the propagation of jets at different scales for individual objects.
When interpreting the results relative to magnetic fields, possible aberrations from relativistic effects have to be considered \citep{2009ApJ...703L.104B,2010ApJ...725..750B}.

The source 1604+159 (J1607+1551) has been classified as Low-spectral peaked (LSP) Quasar with redshift z = 0.4965 \citep{2008ApJS..175..297A}.
It has a Luminosity Distance of $ D_{L}$ = 2799 Mpc and a linear scale of 6.06 kpc/arcsec. 
It was detected by both $\gamma-ray$ detector EGRET \citep{1995ApJS..101..259T} and Fermi \citep{2010ApJS..188..405A,2012ApJS..199...31N}.
It has flat radio spectral and weak radio variability \citep{2014A&A...572A..59M,2022AstBu..77..361S}.
It is a core–jet source, from very long baseline interferometry (VLBI) up to 8.4 GHz VLA-A scales (Jodrell–VLA Astrometric Survey; JVAS; \citet{2006MNRAS.368.1411A}).

Our previous work on the quasar 1604+159 \citep{2024ApJ...965...74H} has studied the evolution of the magnetic field for the core and the jet region up to a distance of $\sim$ 9 mas.
In this study, we further investigate the magnetic field from the parsec to kilo-parsec scale, aiming to gain a deep understanding of the evolution of the magnetic field with distance.
The source has not yet been studied for the magnetic field on a mid to large scale.
The new results indicate that the source carries remarkable information on the magnetic field at different scales.

\section{Observation and data reduction}
\subsection{Observations}
This work contains one observation with the Very Long Baseline Array (VLBA) and three observations with the Very Large Array (VLA), all in dual-polarization mode.

The VLBA project is BH065 observed at 1.7 GHz on 2000 February 07.
The data has 2 intermediate frequencies (IFs) with 16 MHz total bandwidth.
The source was observed for 16 scans, and the on-source time was $\sim$ 64 minutes.

The VLA projects are AS637, AH635, and AH721.
AS637 was conducted at the $L$ band with VLA-B array configuration on 1998 September 21.
It has two spectral windows (spws) with central frequencies of 1.4 and 1.7 GHz and a bandwidth of 50MHz for each.
The source was observed for two scans with an on-source time of 710 seconds.
AH635 was observed at the $X$ and $U$ bands with VLA-C array configuration on 1999 January 29.
The two bands have central frequencies of 8.5 and 22.5 GHz, each containing two spws with 100 MHz total bandwidth.
The source was observed for one scan with an on-source time of 200 and 210 seconds, separately.
AH721 was observed at the $X$ and $U$ bands with VLA-A array configuration on 2000 December 01.
The two bands were made in the same setting as AH635.
The source was observed for one scan with an on-source time of 230 and 240 seconds.
Detailed information on the observation is listed in Table~\ref{tab:observation}.

\subsection{Data Reduction}
\subsubsection{The VLA Data}
The VLA data was downloaded as .exp files.
The .exp files were then converted to .ms files in CASA (Common Astronomy Software
Applications \footnote{\href{https://science.nrao.edu/facilities/vla/data-processing}{https://science.nrao.edu/facilities/vla/data-processing}}).

Before calibration, we selected a reference antenna for each project (listed in Table~\ref{tab:observation}).
We also inspected the data and flagged bad data points for calibrators and the target source.
The ionosphere correction is important for polarimetry and was done for AS637 as it was observed at the $L$ band.
Opacity was corrected for the 22.5 GHz data of AH635 and AH721.

Calibration was done following the pre-upgrade VLA Tutorials provided by NRAO \footnote{\href{https://casaguides.nrao.edu/index.php/Pre-upgrade_VLA_Tutorials}{https://casaguides.nrao.edu/index.php/Pre-upgrade-VLA-Tutorials}}:
\begin{enumerate}
    \item 3C 286 was chosen to be the flux density scale calibrator and the model was set through the task `SETJY' in CASA using the Perley-Butler 2017 standard \citep{2017ApJS..230....7P}. 
    \item After initial gain calibration with the task `GENCAL', the flux density solutions of the target were derived using the task `FLUXSCALE'.
    \item  The old VLA data have single-channel continuum spws, we thus did not solve for R-L delays (right circular and left circular polarization) in our polarization calibration.
    \item The instrumental polarization (D-terms) was solved using the task `POLCAL' with poltype=`D'.
Low-polarized (fractional polarization $< 1\%$) calibrator J1404+286 (OQ 208)  was used for AS637 and 1331+170 for AH635 and AH721.
\item 3C 286 also served as the EVPA calibrator and the polarized parameters were imported manually using the task `SETJY'.
EVPA was corrected using the task `POLCAL' with poltype=`Xf'.
The corresponding calibration error is $\sim 1^{\circ}$.
\end{enumerate}

Each spectral window in the $L$ band data of AS637 was calibrated independently.
After applying the solutions, the source was split from the multisource data sets and converted to the .uvfits files with the task `EXPORTUVFITS'.

\subsubsection{The VLBA Data}
The initial calibration was conducted using the Astronomical Imaging Processing Software (AIPS) package \citep[]{2003ASSL..285..109G} with the standard procedure.

After inspection of the data, Los Alamos (LA) was chosen as the reference antenna for BH065.
The amplitude calibrations were done using the information for all antennas in the gain curve (GC), system temperature (TY), and weather (WX) tables.
The ionospheric delay was corrected via total electron content measurements from global positioning system monitoring.
The phase contributions from the antenna parallactic angles were removed before any other phase corrections were applied.
The delay search, bandpass calibration, and RL delay were performed with the source 1156+295.

The instrumental polarization solutions were solved using the task `LPCAL’. 
Source 0552+398 (DA 193) served as the D-terms calibrator.
According to the results of Monitoring of Jets in Active Galactic Nuclei with VLBA Experiments (MOJAVE) \footnote{\href{https://www.cv.nrao.edu/MOJAVE/sourcepages/0552+398.shtml}{https://www.cv.nrao.edu/MOJAVE/sourcepages/0552+398.shtml}}, DA 193 has a fractional polarization of $\sim 1 \%$ at 15 GHz.
Due to the depolarization effect, the fractional polarization decreases with the decline in frequency.
Thus, at the $L$ band, DA 193 has low fractional polarization ($< 1\%$) and could be treated as an unpolarized source in the task `LPCAL'.
For project BH065, DA 193 was arranged for 2 scans and 8 antennas took part in the observation during the observation.
Solutions of D-terms for antennas Hancock (HN) and Mauna Kea (MK) could not be derived.
However, after applying the solutions to the target source and subtracting the real emission, the RMS for the $Q$ ($\rm 0.15\ mJy\ beam^{-1}$) and $U$ ($\rm 0.15\ mJy\ beam^{-1}$) distributions are close to that for the $I$ distribution  ($\rm 0.13\ mJy\ beam^{-1}$).
In addition, we did not find fake emissions around the source, and the residual maps after subtracting the real ones seem thermal-noise-dominated.
Thus, we believe that the resultant distribution of polarized emission is convincing, as seen in Figure~\ref{fig:BH065}.

We did not correct the EVPA because we did not find any useful information close to the observation date.
However, even if we corrected the offset, the EVPA still could not represent the intrinsic polarization at the $L$ band, and the EVPA needed to continue to be corrected for the RM, which requires multi-frequency information.
Therefore, we only have shown the fractional polarization.
 
\subsection{De-biase and Imaging}
Imaging and self-calibration were carried out in the Difmap package \citep{1997ASPC..125...77S}.
The \textit{I}, \textit{Q}, and \textit{U} distributions were then obtained using the fully self-calibrated visibilities with natural weighting.

According to the study of \citet{2023MNRAS.520.6053P}, the total and linear polarization intensity
distributions suffer the CLEANing bias, which comes from the residual (i.e. uncleaned) components.
This bias could be solved by deep cleaning \citep{1995AAS...18711202B,2023MNRAS.520.6053P}.
After subtracting the real emissions identified visually, we put boxes covering the whole structure of the source and made a deeper clean down into the noise level.

The distributions of the polarization intensity ($P = \sqrt{Q^2+U^2}$), polarization fraction (i.e. $m=p/I$), and EVPA ($\frac{1}{2}\mathrm{tan}^{-1}(\frac{U}{Q})$) were calculated pixel by pixel. 

During the propagation from extra-galactic astrophysical objects to Earth, electromagnetic waves suffer the Faraday effect from our Milky Way \citep{2022A&A...657A..43H}.
It is expected that the Faraday effect at angular scales of arc minutes or larger is mostly attributed to the local effects \citep{2010ApJ...723..476A}.
Our target has Galactic coordinates $(l,b)=(29.378427^{\circ}, 43.409096^{\circ})$ \footnote{\href{https://ned.ipac.caltech.edu/byname}{https://ned.ipac.caltech.edu/byname}}.
Based on the up-to-date RM catalogs \footnote{\href{https://github.com/CIRADA-Tools/RMTable}{https://github.com/CIRADA-Tools/RMTable}}, we found three sources with $28^{\circ}<l<30^{\circ}$ and $42^{\circ}<b<45^{\circ}$, listed in Table~\ref{tab:Milky Way RM}.
We took the average RM measured for these three sources as the RM from our Milky Way to the target, which is 20.4 $\rm rad \ m^{-2}$.
The EVPA distributions of all the frequencies were then corrected for this effect before further analyses (The $L$ band data is affected the most, while the impact on 8.5 and 22.5 GHz data is negligible.)

According to \citet{1974ApJ...194..249W}, the polarized intensity, $p$, obeys a Rayleigh distribution and there exists a Ricean bias between the observed polarized intensity and the best estimate of the true polarization.
Following \citet{1974ApJ...194..249W,2023MNRAS.520.6053P,2023MNRAS.523.3615Z}, we corrected the bias by calculating the true values $p_{\mathrm{true}} \sim p(1-(\frac{\sigma_{p}}{p})^{2})^{\frac{1}{2}}$.

The uncertainties of $\sigma_{I}$, $\sigma_{Q}$, and $\sigma_{U}$, were obtained using the approach of \citet{2012AJ....144..105H}, considering the effect of the D-terms for each pixel.

The uncertainties of $p$, $m$, and EVPA are
\begin{equation}
\begin{aligned}
    &\ \sigma_{p}=\frac{\sigma_{Q}+\sigma_{U}}{2} \\
    &\ \sigma_{m}=m((\frac{\sigma_{p}}{p})^{2} + (\frac{\sigma_{I}}{I})^{2} + 0.05^{2})^{\frac{1}{2}} \\
    &\ \sigma_{\mathrm{EVPA}}=((\frac{\sigma_{p}}{2p})^{2} + (1^{\circ}\times\frac{\pi}{180^{\circ}})^{2})^{\frac{1}{2}},
\end{aligned}
\end{equation}
where, in $\sigma_{m}$ we added in quadrature the antenna gains amplitude uncertainty of 0.05 \citep{2023MNRAS.520.6053P}, and in $\sigma_{\mathrm{EVPA}}$ the calibration error of $1^{\circ}$ \citep{2012AJ....144..105H}.

Based on the derived uncertainties, we clipped pixels with $I<3\sigma_{I}$, $p<3\sigma_{p}$, and $m<3\sigma_{m}$ for the polarization distributions (i.e. $p$, $m$, and EVPA).

\begin{table}
    \centering
    \caption{RM from our Milky Way}
    \begin{tabular}{c c c c c}
         \hline
         \hline
         Ra& Dec& $l$ &$b$ &RM\\
         ($^\circ$)&($^\circ$)&($^\circ$)&($^\circ$)& rad ($\rm m^{-2}$)\\
         \hline
         240.351537&  15.616996&  28.3305&  44.5805&  15.7 \\
         240.656328&  15.737358&  28.6455&  44.3568&  21.1 \\
         240.934037&  15.795774&  28.8649&  44.1328&  24.5 \\
         \hline
    \end{tabular}
    \label{tab:Milky Way RM}
\end{table}

\begin{table*}
    \centering
    \caption{Observation information}
    \begin{tabular}{c c c c c c c c c}
         \hline
         \hline
         Epoch & Project ID & Array & Frequency & Bandwidth & On Source time & D-term & EVPA & Reference Antenna \\
           &&&&&&Calibrator& Calibrator&\\
               & & & (GHz) & (MHz) & (second) & &&\\
         \hline
          2000-Feb-07 &BH065 &VLBA & 1.7  & 16 & $64\times60$ & DA 193 && LA \\
         \hline
          2000-Dec-01 & AH721 & VLA-A & 8.5  & 100 & 230 & 1331+170 & 3C 286& VA06\\
          & & & 22.5 &  & 240 & & \\
         \hline
         1999-Jan-29 & AH635 & VLA-C & 8.5  & 100 & 200 & 1331+170 & 3C 286&VA07\\
         & & & 22.5 &  & 210 & & \\
         \hline
         1998-Sep-21 & AS637 & VLA-B & 1.4  & 50 & 710 & OQ 208 & 3C 286&VA28\\
         & & & 1.7 &  & & & \\
         \hline
    \end{tabular}
    \label{tab:observation}
\end{table*}

\begin{table*}
    \centering
    \caption{Image statistics summary}
    \begin{tabular}{c c c c c c}
        \hline
        \hline
         Epoch & Project ID & Frequency & Resolution & $\sigma$ & Integrated Flux Density \\
          & & (GHz) & (MilliArc second) & (Jy $\rm beam^{-1}$) & (Jy)\\
          \hline
          2000-Feb-07 &BH065& 1.7& $5.03 \times 9.87$ (-4.59) & $1.3 \times 10^{-4}$ & $0.34\pm0.03$\\
          \hline
         2000-Dec-01 & AH721 & 8.5 & $217.54 \times 250.23$ (5.48)&$8.3\times 10^{-5}$& $0.51\pm0.05$ \\
          & & 22.5 & $82.14 \times 95.97$ (5.91) &$2.1 \times 10^{-4}$ & $0.52\pm0.05$\\
        \hline
        1999-Jan-29 & AH635 & 8.5& $2344.24 \times 2818.22$ (-21.36) &$8.5 \times 10^{-5}$  & $0.56\pm0.06$\\
        & & 22.5 & $923.90 \times 1042.75$ (-16.36) &$3.7 \times 10^{-4}$  & $0.53\pm0.05$\\
        \hline
        1998-Sep-21 & AS637 & 1.5& $4224.03 \times 5280.15$ (49.61) &$1.5 \times 10^{-4}$  & $0.62\pm0.06$\\
        \hline
    \end{tabular}
    \label{tab:Image stat.}
\end{table*}

\begin{table*}
    \centering
    \caption{Model fitting parameters. Columns: (1) observation date (2) model component, (3) frequency, (4) total flux density, (5) model distance, (6) model position angle, (7) model major axes, (8) model axial ratio, (9) direction of major axes, (10)flux density ratio between EH and WH.}
    \begin{tabular}{c c c c c c c c c c}
        \hline
        \hline
         Epoch & Component & Frequency & $ S_{\mathrm{tot}}$ & $r$ & $\theta$ & Major axes & Axial ratio & Phi & Ratio \\
         & & (GHz) & (mJy) & (mas) & ($^{\circ}$) & (mas) && ($^{\circ}$)&\\
          (1) & (2) & (3) & (4) & (5) & (6) & (7) & (8) & (9) & (10)\\
         \hline
         1999-Jan-29 & EH& 8.5 & 8.5 & 10918.9 & 96.0 & 1738.7 & 0.40 & -57.3 & 2.9\\ 
         & WH& & 2.9 & 11452.0 & -93.0 & 2727.6 & 0.33 & -44.0& \\
         \hline
        1998-Sep-21 & EH& 1.4 & 41.8 & 10413.3 & 97.3 & 2687.1 & 0.70 & -68.5& 2.3\\ 
         & WH& & 17.9 & 11761.0 & -92.5 & 3138.3 & 0.78 & -15.3 & \\
         1998-Sep-21 & EH& 1.7 &35.7&10468.4& 97.2& 2793.0 & 0.60&-70.4& 1.7 \\ 
         & WH& &20.5 & 11380.7 & -93.0 & 4199.5 & 0.90 & 12.0& \\     

         \hline
    \end{tabular}
    \label{tab:Model fitting}
\end{table*}

\section{results}
In this section, we provide the distributions of the total intensity, polarized intensity, and fractional polarization, for the source 1604+159 from parsec to kilo-parsec scale.
For the largest structure detected, we also analyzed the RM, RM-corrected EVPAs, and spectral index distributions.
As the jet of the source has low RM values ($<200$ rad $\rm m^{-1}$)\citep{2024ApJ...965...74H}, the observed polarization with frequency $>$  4 GHz could represent the intrinsic one, which has a perpendicular relation with the magnetic field projected on the plane of the sky in the optically thin regions \citep{1970ranp.book.....P}.
The results show that the source carries substantial information on the magnetic field at hundreds of mas, arcsec, and tens of arcsecond scales.
Table~\ref{tab:Image stat.} lists the information on the images.

\subsection{Inner Jet}
\begin{figure}[htbp]
    \centering
    \includegraphics[width=\linewidth]{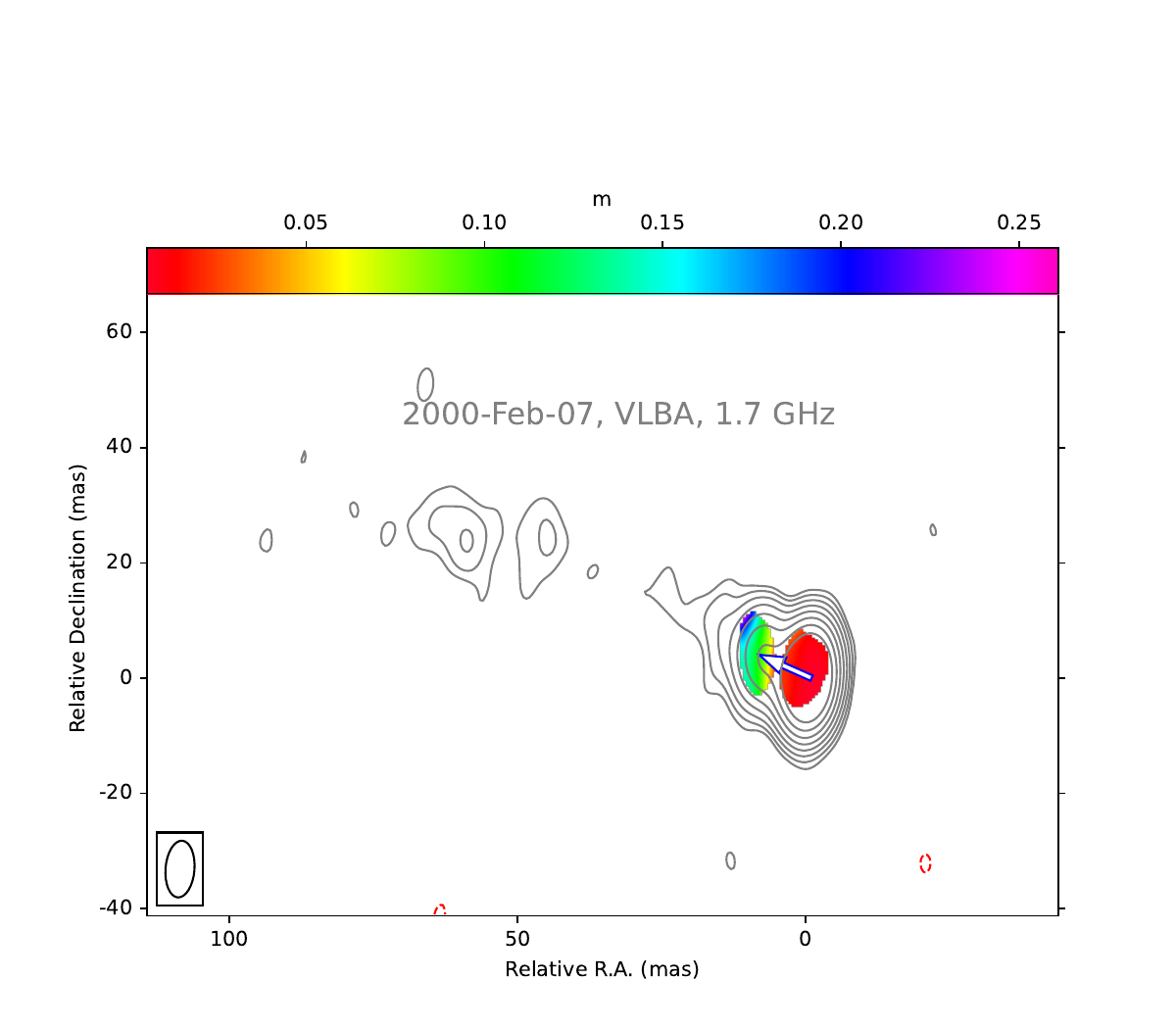}
    \caption{Distribution of fractional polarization (color) at 1.7 GHz for the observation on 2000 February 07 superimposed on the corresponding total intensity contours.
    Contours are $3\sigma \times (-1, 1, 2, 4, 8, 16, 32, 64, 128)$.
    The blue arrow indicates the jet direction at the parsec scale.
    }
    \label{fig:BH065}
\end{figure}

Figure~\ref{fig:BH065} plots the distributions of total intensity and fractional polarization at 1.7 GHz observed on 2000 February 07.
Compared with the result at 4.6 GHz \citep{2024ApJ...965...74H}, which has shown jet structure to the 25 mas distance from the core, the jet at 1.7 GHz extends to the $\sim$ 90 mas distance.
The jet starts to change its direction at the place of $\sim$ 50 mas.

Polarization is detected at the core and the place of $\sim$ 10 mas, where a standing shock is located for $\sim$ 20 years \citep{2024ApJ...965...74H}.
The fractional polarization is $\sim 15 \%$, higher than that of the core, and consistent with the results detected in the previous work \citep{2024ApJ...965...74H}.

\subsection{Intermediate Jet}
\begin{figure*}[htbp]
    \centering
    \includegraphics[width=\linewidth]{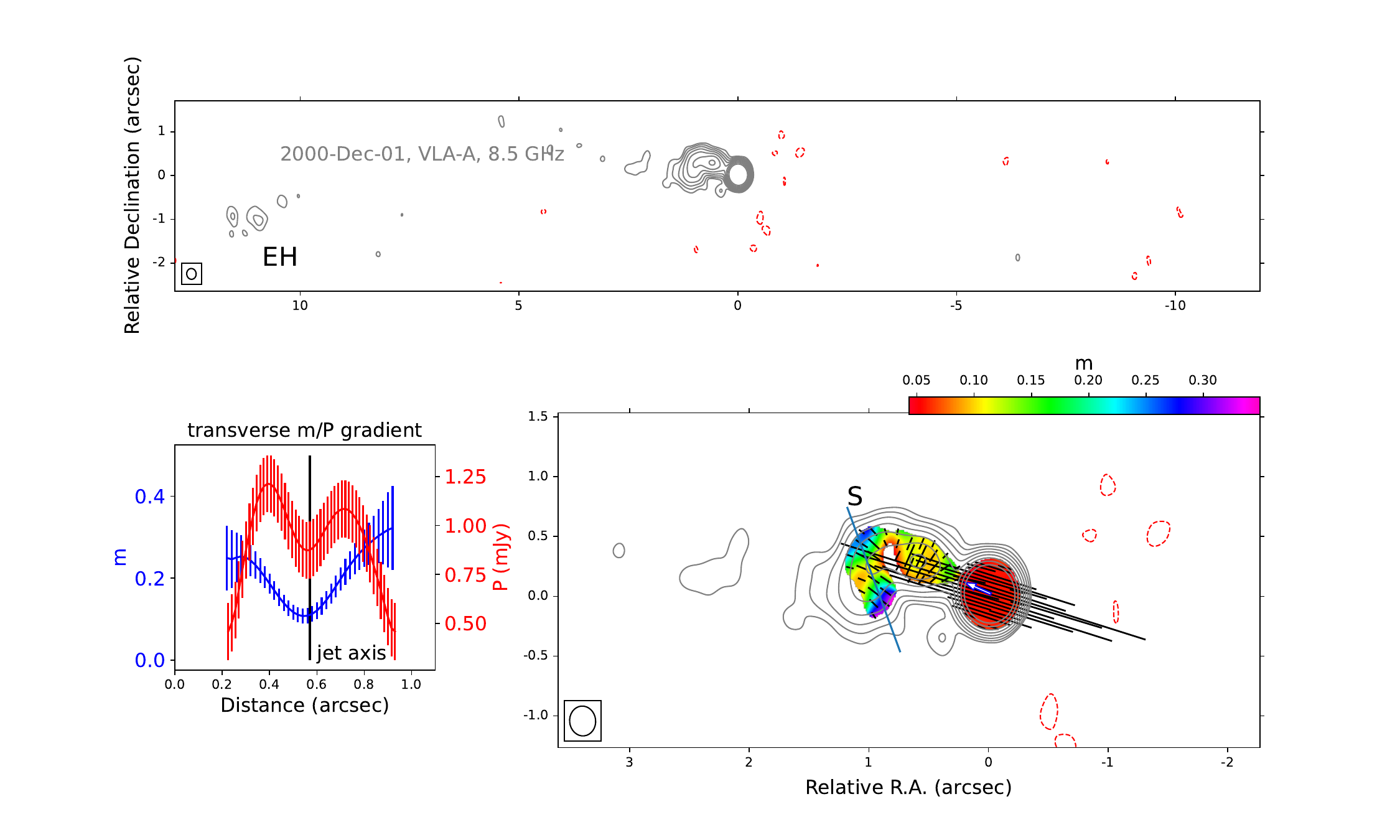}
    \includegraphics[width=0.6\linewidth]{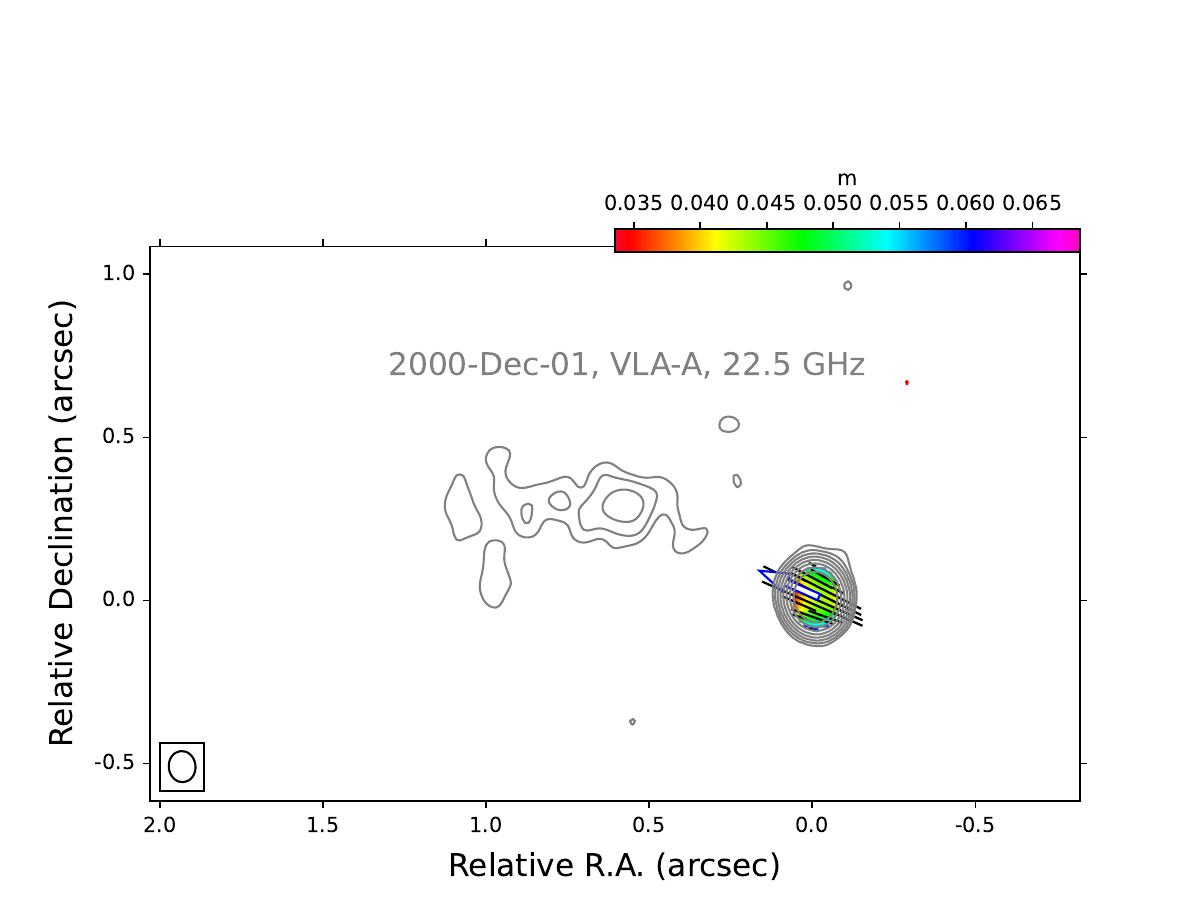}
    \caption{Distributions of linear polarization (sticks) and fractional polarization (color) at 8.5 and 22.5 GHz for the observation on 2000 December 01 superimposed on the corresponding total intensity contours.
    Contours are $3\sigma \times (-1, 1, 2, 4, 8, 16, 32, 64, 128)$.
    The blue arrow indicates the jet direction at the parsec scale.
    There is a hotspot (EH) located on the eastern side, which has been detected at two other VLA observations (Figures~\ref{fig:AH635} and \ref{fig:AS637 RM}).
    There exists a Western counterpart, which is not detected at this epoch due to the limited sensitivity.
    Distributions of fractional polarization and polarized intensity along a slice across the curved jet indicate the enhancement in the magnetic field at the jet edges.
    }
    \label{fig:AH721}
\end{figure*}

Figure~\ref{fig:AH721} plots the distributions of $I$, $p$, and $m$ at 8.5 and 22.5 GHz observed in 2000 December 01 with the VLA-A array.

At 8.5 GHz, the upper panel of the plot shows a jet structure and an eastern hotspot (EH) located at the distance of $\sim$ 11 mas, which is confirmed in the results of the two other observations (Figures~\ref{fig:AH635} and \ref{fig:AS637 Lband}).
There exists a western hotspot (Figures~\ref{fig:AH635} and \ref{fig:AS637 Lband}) at the opposite of EH, similar distance relative to the core.
In Figure~\ref{fig:AH721}, we did not detect the counterpart probably due to the limited sensitivity.

The jet extends to a distance of $\sim$ 2 arcsec.
This structure is consistent with the result of \citet{1998MNRAS.293..257B}, which detected highly similar emissions in the same region.
At $\sim$ 0.5 arcsec, there is a knot, beyond which the jet starts to bend to the south.
Structures of the knot and the bend are apparent at 22.5 GHz (lower panel of Figure~\ref{fig:AH721}).
The map shows relatively weak and extended emissions between $1\sim3$ mas, the eastern side of the bend (upper panel of Figure~\ref{fig:AH721}).
This may hint that the jet bends up and propagates to the east after a bend to the south.
Future high-sensitivity observation of VLA \citep{2011ApJ...739L...1P} is needed to confirm this extended structure.

The polarization of the VLA core orients along the jet direction for both frequencies.
However, the jet at 8.5 GHz shows polarization perpendicular to the local jet direction, which is quite different from the oblique polarization distribution at $\sim$ 3.7 and 9 mas and longitudinal distribution at the upper jet region \citep{2024ApJ...965...74H}.
Perpendicular polarization indicates a longitudinal magnetic field structure.

Before the jet bends to the south, the polarization close to the jet axis has strong emission and decreases to the edges.
As the jet bends, the polarization at the edges becomes stronger.
We extracted a slice across the jet bend and plotted the distributions of polarized intensity and fractional polarization along the slice (Figure~\ref{fig:AH721}).
As seen, the values of polarized intensity first increase and then decrease from the jet axis to the edges.
The values of fractional polarization keep increasing.
Higher fractional polarization at the jet edge when bending is also observed for sources J0738+1742 \citep{2012evn..confE..94H} and Mrk 501 \citep{2005MNRAS.356..859P}.
In addition, J0738+1742 and Mrk 501 show perpendicular polarization at the place with high fractional polarization.
A comprehensive interpretation of this will be discussed in the discussion section.

\subsection{Outer Jet}
\begin{figure*}[htbp]
    \centering
    \includegraphics[width=0.8\linewidth]{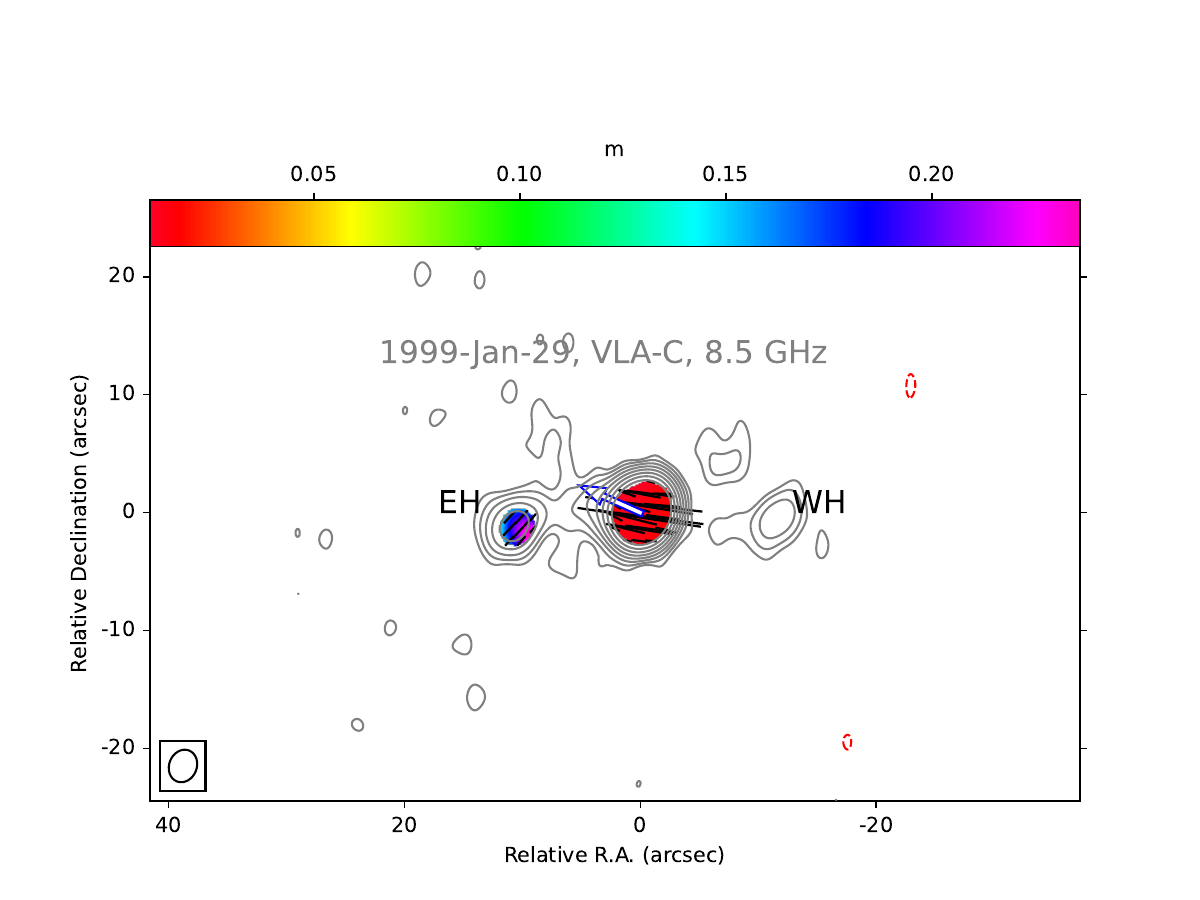}
    \includegraphics[width=0.8\linewidth]{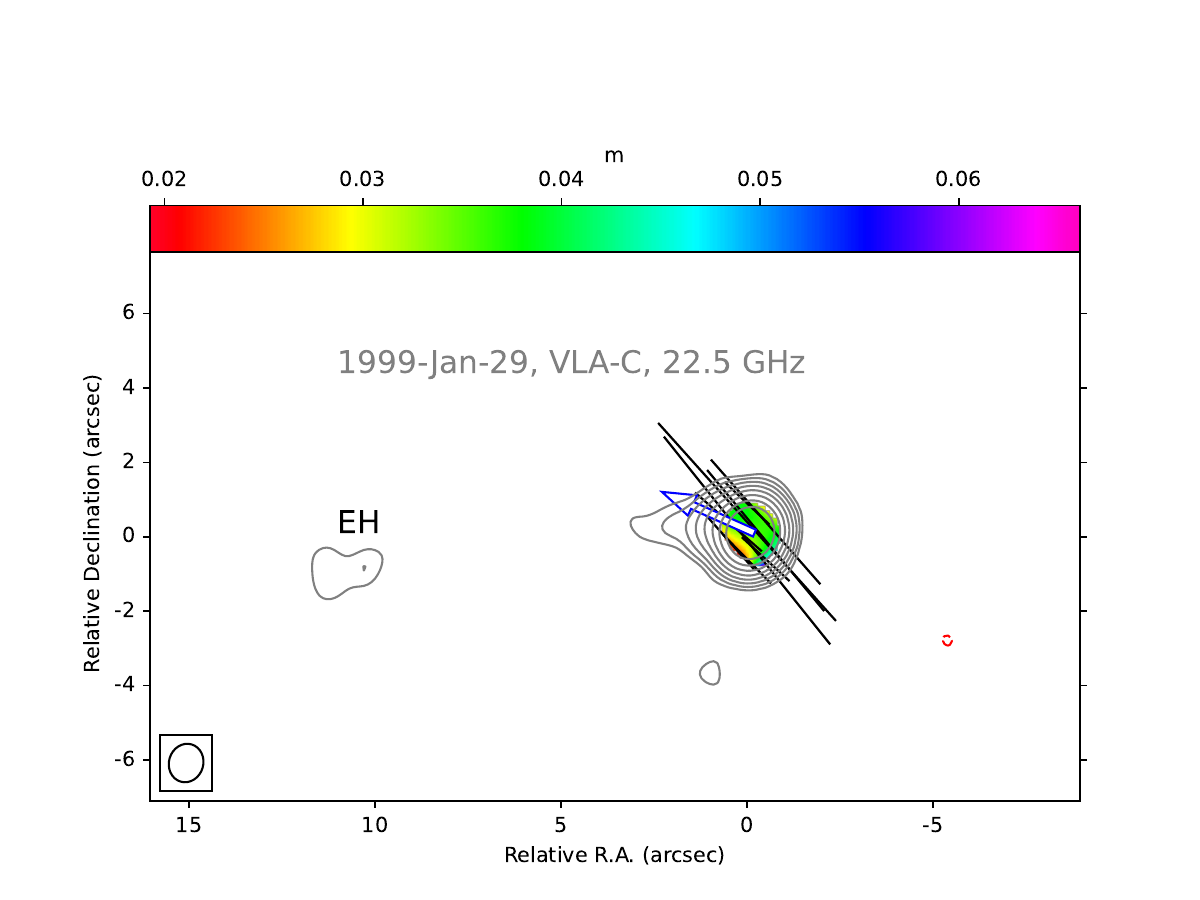}
    \caption{Distributions of linear polarization (sticks) and fractional polarization (color) at 8.5 and 22.5 GHz for the observation on 1999 January 29 superimposed on the corresponding total intensity contours.
    Contours are $3\sigma \times (-1, 1, 2, 4, 8, 16, 32, 64, 128)$.
    The blue arrow indicates the jet direction at the parsec scale.
    Two hotspots are located on the eastern and western sides, labeled as `EH' and `WH'.
    The diffuse emissions around EH at 8.5 GHz come from the eastern lobe as plotted in Figure~\ref{fig:AS637 Lband}.}
    \label{fig:AH635}
\end{figure*}

Figure~\ref{fig:AH635} plots the distributions of $I$, $p$, and $m$ at 8.5 and 22.5 GHz observed on 1999 January 29 with the VLA-C array.

As can be seen in the 8.5 GHz map, after launch, the eastern jet propagates as a collimating structure and terminates at the east hotspot (bright, compact shocks), confirming the detection on 2000 December 01 (Figure~\ref{fig:AH721}).
According to model fitting results, the hotspot is located at a distance of 10.9 arcsec from the core with a direction of $96_{.}^{\circ}0$.
The measured flux density, $S_{\rm est}$, is 8.5 mJy.

Polarization emission is detected in the eastern hotspot, showing a large separation from the jet direction.  
As will be shown in the following section and Figure~\ref{fig:AS637 RM}, the RM around the hotspot is small enough ($< 25 \ \rm rad \ m^{-2}$) that the observed polarization in the hotspot could represent the intrinsic polarized direction, indicating an oblique magnetic field structure.

On the western side, a hotspot (WH) is also detected.
According to model fitting results, the hotspot is located at a distance of 11.5 arcsec from the core with a direction of $-93_{.}^{\circ}0$, a similar position symmetrical with the eastern one.
The measured flux density, $S_{\rm wes}$, is 2.9 mJy.
$S_{\rm est}$/$S_{\rm wes}$=2.9, suggesting the effect of Doppler beaming still exists at a large scale.
The polarization is not detected at the western hotspot due to the Doppler beaming effect and the non-enough sensitivity.

The jets propagate along the east-west direction, different from the direction at the parsec scale, indicating a change of direction during the propagation, to which the bends at $\sim$ 50 mas (Figure~\ref{fig:BH065}) and 1 $\sim$ 2 arcsec contribute (Figure~\ref{fig:AH721}).

The polarization of the core at 8.5 GHz in general orients along the jet direction of the parsec scale, the same as the result at 8.5 GHz on 2000 December 01 (Figure~\ref{fig:AH721}).
However, at 22.5 GHz, the core shows obvious polarization variation with time.
On 1999 January 29, the polarization of the core had a direction of $\sim 45^{\circ}$, while on 2000 December 01 the direction changed to $\sim 66^{\circ}$.

\subsection{Emission at a Larger Scale}
\begin{figure*}[htbp]
    \centering
    \includegraphics[width=0.6\linewidth]{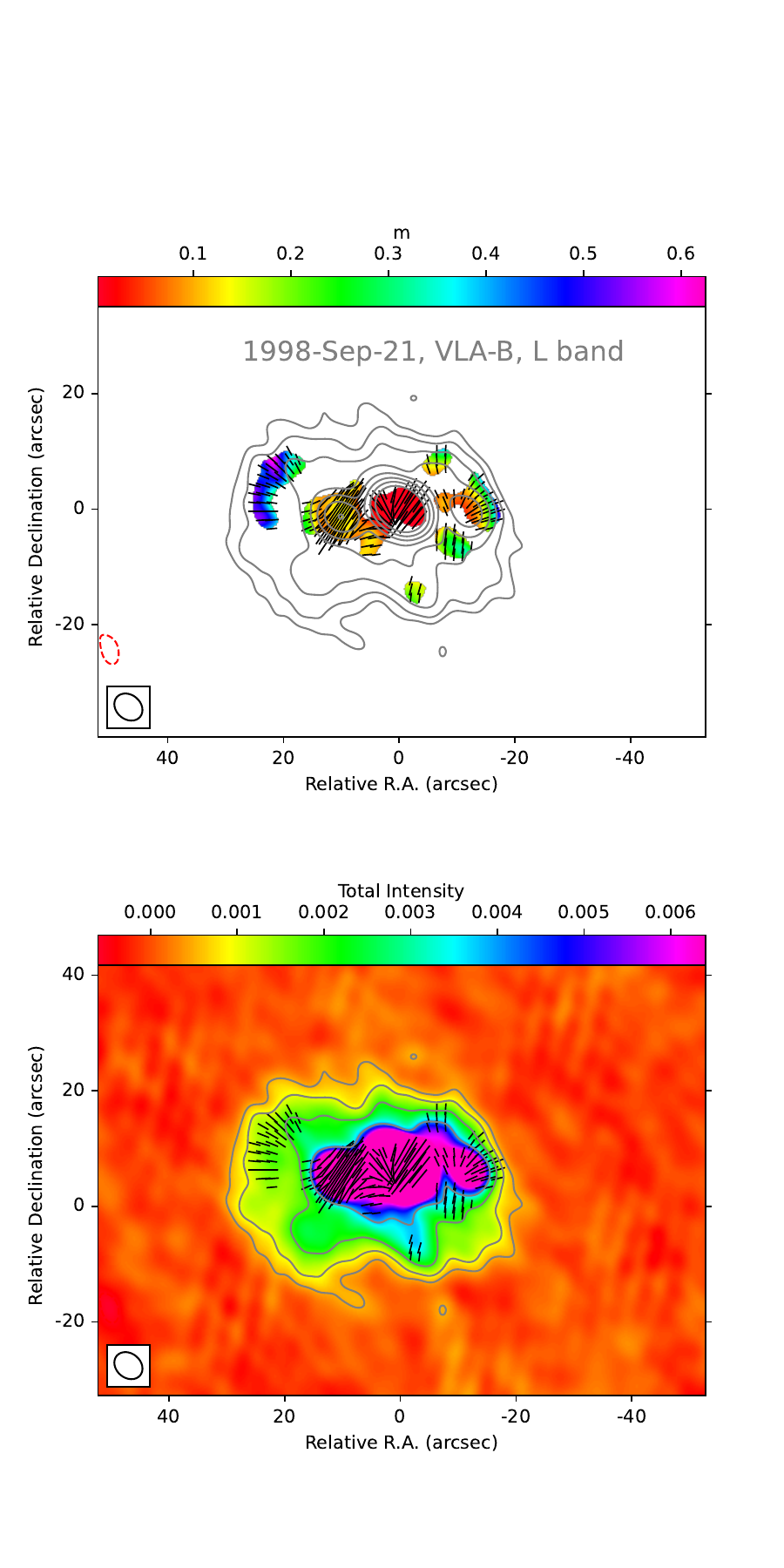}
    \caption{Distributions of linear polarization (sticks), fractional polarization (color in upper panel), and total intensity (color in bottom panel) at the $L$ band for the observation on 1998 September 21 superimposed on the corresponding total intensity contours.
    To show the extended emissions, the pixels with values $>$ 0.15 $\times$ peak intensity were replaced with 0.15 $\times$ peak intensity in the bottom plot.
    Contours are $3\sigma \times (-1, 1, 2, 4, 8, 16, 32, 64, 128)$.}
    \label{fig:AS637 Lband}
\end{figure*}

Figure~\ref{fig:AS637 Lband} shows the results at the $L$ band, which is the combination of the two spws centered at 1.4 and 1.7 GHz to obtain better sensitivity, observed on 1998 September 21 with the VLA-B configuration.
Unlike the 8.5 GHz observation on 1999 January 29 (Figure~\ref{fig:AH635}), which detects only the bright, compact core and hotspots, the results at the $L$ band show more extensive structures including the emission of several lobes.

To show the extended emissions in detail, we also plotted the color distribution for the total intensity and marked the stronger emissions from hotspots, jets, and the core by replacing the pixels having values $>$ 0.15 $\times$ peak intensity with 0.15 $\times$ peak intensity.
This quasar has a cocoon-like shape with several bulges (shown by the green and blue color in the bottom panel of Figure~\ref{fig:AH635}) surrounding the hotspots, jets, and core.
The fractional polarization is generally higher with further distance from the core.

We fitted the Gaussian models for EH and WH for both the 1.4 and 1.7 GHz data and summarised them in Table~\ref{tab:Model fitting}.
The distance and position angle of EH and WH at this epoch are consistent with the result in 1999.

We obtained the spectral index, RM, and RM-corrected EVPA distributions for the target source 1604+159 to analyze its opacity and magnetic field with the two-frequency data.
The (u,v) coverage was ﬁrst ﬂagged to the same range in which the minimum UV distance is the smallest value of the highest frequency and the maximum is the largest value of the lowest
frequency.
Stokes maps were then produced with the same map size and cell size and restored to the lowest-frequency beam size.
Finally, maps were aligned by shifting the map centers to the position of EH.
Maps were clipped pixels with $I<3\sigma_{I}$, $p<3\sigma_{p}$, and $m<3\sigma_{m}$ for the polarization distributions (i.e. $p$, $m$, and EVPA).

The RM and RM-corrected EVPA distributions were derived from the equation \citep{1975clel.book.....J},
\begin{equation}
    \label{RM}
    \chi_{\rm obs} - \chi_{\rm o} = \frac{e^3\lambda^2}{8\pi^2\epsilon_om^2c^3} \int n_e\textbf{\textit{B}}\cdot d\textbf{\textit{l}} \equiv \rm RM\lambda^2,
\end{equation}
where $\chi_{\rm obs}$ is the observed polarization angle, $\chi_{\rm o}$ is the intrinsic polarization angle, $\lambda$ is the wavelength, $n_{e}$ is the density of the electrons in the medium, and $\textbf{\textit{B}}\cdot d\textbf{\textit{l}}$ is the projected $\textit{\textbf{B}}$ filed component on the line of sight (LOS) multiplied by the integral element $d\textit{l}$.

The rotation of EVPA at the $L$ band could be large even if the RM has a magnitude of tens of rad $\rm m^{-2}$.
Thus, unlike higher frequencies, EVPA must be corrected to represent the intrinsic polarization direction and the magnetic field structure.

\begin{figure*}
    \centering
    \includegraphics[width=\linewidth]{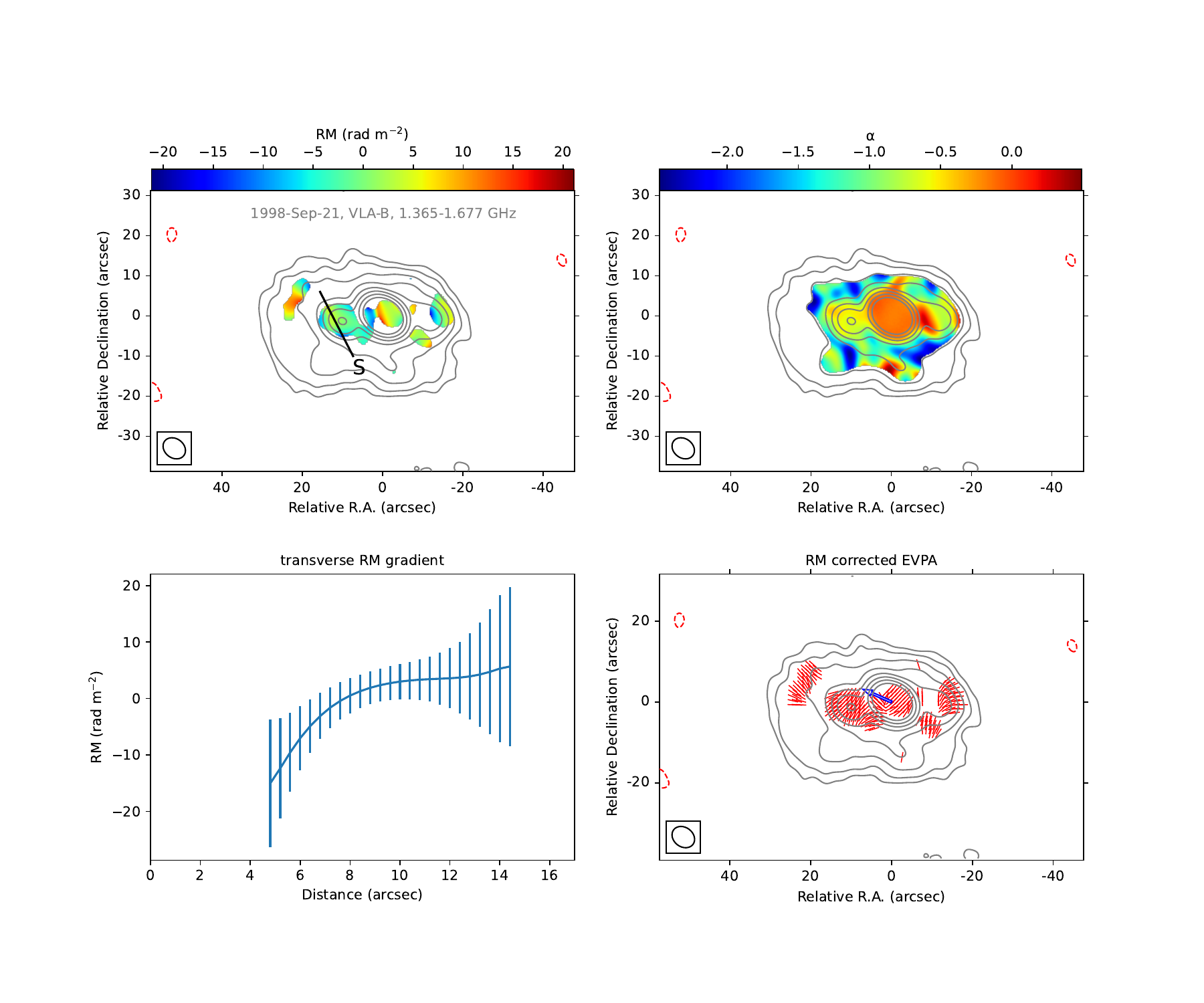}
    \caption{Distributions of RM, RM-corrected EVPA, and spectral index for the observation on 1998 September 21 superimposed on the 1.7 GHz total intensity contours.
    Contours are $3\sigma \times (-1, 1, 2, 4, 8, 16, 32, 64, 128)$.
    The blue arrow indicates the jet direction at the parsec scale.
    The bottom left panel plots a transverse RM gradient, in which the start point `S' is labeled in the RM map.}
    \label{fig:AS637 RM}
\end{figure*}

The spectral index distribution was obtained using $S \propto \nu^{\alpha}$.
When calculating the spectral index and its uncertainty, we added a calibration uncertainty of 0.05 in quadrature.
Pixels were clipped when the total intensity level was less than $3\sigma_\nu$ at the corresponding frequency, where
\begin{equation}
    \rm \sigma_{\nu} = \sqrt{\sigma^2_{rms} + (1.5\sigma_{rms})^2} \approx 1.8\sigma_{rms},
\end{equation}
where $\sigma_{\rm rms}$ is the thermal noise of the image taken at a corner of each map \citep{2014AJ....147..143H}.
We also blacked the pixels with uncertainty of spectral index $>1$.
The resultant map is shown in Figure~\ref{fig:AS637 RM}.

Due to the optically thin and steep spectral nature (Figure~\ref{fig:AS637 RM}), the polarization emission in the western hotspot at the $L$ band is detected, which shows a symmetrical distribution with the eastern emission relative to the core.
The polarization in the center of the eastern hotspot has the same direction as the result of the 8.5 GHz data observed on 1999 January 29 (Figure~\ref{fig:AH635}), indicating a stable oblique polarization and magnetic field structure with time.
At the outer region of the eastern hotspot, the polarization changes smoothly to be normal on the edge, indicating the magnetic field orients along the outer edge of the hotspot, which also existed in other hotspots of quasars including PG 0003+158, PG 0007+106, PG 1048-090, PG 1100+772, PG 1103-006, and 3C 273 \citep{2023MNRAS.519.2773B}.

The polarization of the core at the $L$ band shows a large separation from the jet direction, quite different from the results at 8.5 GHz on the two other observations (Figures~\ref{fig:AH721} and \ref{fig:AH635}).
The core has flat spectral index values (Figure~\ref{fig:AS637 RM}).
This suggests that the emission is dominated by the optically thick region and/ or results from the blending of emissions from the VLBI core and the jet region.
Due to the steep spectral nature, the polarized emission is stronger at the $L$ band, compared with higher frequencies.

Interestingly, for the RM covering the eastern hotspot, the northern part has an average value larger than the southern part.
We then extracted a transverse slice and obtained a RM gradient.
The gradient shows a sign change along the slice.
In the work of \citet{2024ApJ...965...74H}, we have found RM gradients in two epochs in the core region of 1604+159, as strong evidence of the existence of a helical magnetic field.
The RM gradient detected in the eastern hotspot region may suggest the global helical magnetic field may exist at a large scale.
Transverse RM gradients are also detected at kilo-parsec scale for sources including 5C 4.114 \citep{2015A&A...583A..96G} and Coma A \citep{2017Galax...5...61K}, some of which are found in the regions of hotspots.
To note, as the VLA has been upgraded to capable of wideband observation \citep{2011ApJ...739L...1P}, leading to higher sensitivity, future observation on this source would hopefully confirm this gradient with a higher significance level.

\section{discussion}
In previous work \citep{2024ApJ...965...74H}, we investigated the polarization distribution for 1604+159 from the VLBI core region to the jet region of $\sim$ 9 mas.
We have found evidence that 1604+159 has a global helical magnetic field extending from the jet base to the outer jet region.
The helical magnetic field has a large intrinsic pitch angle ($\sim$ $80^{\circ}$) in the jet rest frame and the observed polarization direction depends on the viewing angle \citep{2013MNRAS.430.1504M}.

With distance, the helical magnetic field configuration would become more and more toroidal-dominated if there are no affecting factors including the interaction with ambient media \citep{2021ApJ...906..105C}.
If the jet does not obviously change its viewing angle when propagating, the observed polarization direction is expected to be parallel to the jet direction.
As the jet propagates, the effect of the environment on the jet may lead to the jet morphology changing beyond expected and the local magnetic field could be changed to deviate from the global structure. 

In the jet region of 1604+159, several bright, compact components (located at $\sim$ 3.7 and 9 mas from the core) show oblique polarization distributions, suggesting oblique magnetic field structures on the plane of the sky.
Evidence indicates the nature of shocks for these components, breaking the structure of the global helical magnetic field.

In this section, we further probe the magnetic field for the source to further distances (up to $\sim$300 kilo-parsec).
The source shows a symmetrical Fanaroff-Riley-Class-I-like structure (Figure~\ref{fig:AS637 Lband}), with the core occupying $0.45/0.62=73\%$ of the total integrated flux density at the $L$ band.

\subsection{Magnetic Field in the Core}
Generally, the core (total and polarized intensity) detected by VLA comprises the VLBI core and part of the jet at the parsec scale, depending on the resolution.
For the core at 8.5 GHz on 2000 December 01 (Figure~\ref{fig:AH721}), the polarized emissions could come from blending the VLBI core and the inner jet (e.g. the oblique polarization).
For the core at 8.5 GHz on 1999 January 29 (Figure~\ref{fig:AH635}), the polarized emissions also blend with the emissions from the intermediate jet (Figure~\ref{fig:AH721}).
Interestingly, the polarization of the VLA core at 8.5 GHz in general orients along the jet direction of the parsec scale at the two epochs.

In the previous study \citep{2024ApJ...965...74H}, we have found the polarization in the VLBI core is much stronger than in the jet.
Moreover, the blending is the outcome of the vector sum for the polarized emission.
Therefore, the polarization detected in the core by VLA is dominated by the VLBI core at 8.5 GHz, which means the variation in the VLA core may reflect the variation in the VLBI core.

At 8.5 GHz and lower frequencies, the polarization of the VLBI core behaves stable with time \citep{2024ApJ...965...74H}, which orients along the jet direction at parsec scale.
Integration of the much stronger VLBI core and weaker jet gives the parallel polarization direction of the VLA core, even though adding the transverse polarization (Figure~\ref{fig:AH721}).

Due to the steep-spectral nature of the jet and limited sensitivity, it's the same as the polarization distribution at higher frequencies (e.g. 22.5 GHz).
At 22.5 GHz, the VLA core shows obvious polarization variation with time.
On 1999 January 29, the polarization of the core had a direction of $\sim 45^{\circ}$ (Figure~\ref{fig:AH635}), while on 2000-Dec-01 the direction changed to $\sim 66^{\circ}$ (Figure~\ref{fig:AH721}).
This is because the VLBI core with the higher frequency is located closer to the central engine (e.g. the supermassive black hole and/or the accretion disk), which has a more complex physical condition, leading to the polarization evolution at 22.5 GHz core.
The significant variation is consistent with the previous work \citep{2024ApJ...965...74H}, in which the VLBI core at a higher frequency (i.e. 15 GHz) shows a significant polarization variation with time.
The evolution of magnetic field structure, strength, and electron density of the surrounding medium are possible origins.

In terms of the variation of the magnetic field, “turbulent extreme multi-zone” (TEMZ) provides a possible deeper origin \citep{1985ApJ...298..114M,2004ApJ...613..725S,2005ApJ...629...52S,2014ApJ...780...87M,2017Galax...5...63M}.
Observations have shown that the observed core of some sources is a standing shock \citep{2008ASPC..386..437M,2008Natur.452..966M}, giving a stable transverse magnetic field structure.
If this is the case, at the jet base, a jet feature as turbulence with disordered and ordered structures (e.g. helical magnetic field) passes through the standing shock region, causing the observed $B$-field variation in the radio core.
This needs long-term monitoring at millimeter and optical bands, as conducted by \citet{2008Natur.452..966M}
for the blazar BL Lacertae.
Higher frequency observation points to a closer position to the central region and has a higher resolution capable of showing the core in more detail \citep{2021ApJ...910L..12E,2024ApJ...964L..25E}.

\subsection{Magnetic Field in the Intermediate Jet}
At a distance of $\sim$ 1 arcsec, the jet shows a complex structure (Figure~\ref{fig:AH721}).
It comprises a straight collimating jet region and a curved jet region.
Generally, the emission shows transverse polarization, quite different from the milli-arcsec scale, at which the polarization is oblique.
The polarization distributions in the two regions are also different, indicating different physical conditions.

Before bending to the south (Figure~\ref{fig:AH721}), the transverse polarization gradient suggests that the projected magnetic field is parallel to the local jet.
The emission at the jet axis is strong and becomes weak towards the edges.

Under the global helical magnetic field assumption, according to the analytical solution of \citet{2021ApJ...906..105C} for magnetically dominated astrophysical jets, the helical magnetic field would become more and more toroidal-dominated, i.e. the intrinsic pitch angle naturally becomes more and more large with the ideal condition unaffected by outer factor.
However, in practice, the pitch angle of the helical magnetic field in the jet partly depends on the ratio of the rotation and outflow velocities \citep{2021Galax...9...58G}.
Thus, if the jet is accelerated during propagation, the pitch angle would become small, leading to the magnetic field along with the local jet direction and transverse polarization detected \citep{2002PASJ...54L..39A,2000MNRAS.319.1109G}.
According to the study of \citet{2013MNRAS.430.1504M}, the transverse polarization could arise when the intrinsic helix pitch angle is small ($\gamma^{'}\sim10^{\circ}$) in the rest frame of the jet.

The perpendicular polarization could also result from the velocity shear \citep{2014MNRAS.437.3405L}.
During the propagation, the jet passes through the surrounding medium, forming the `spine-sheath' structure, with faster components at the jet axis and slower components at the shear layers.

As the jet bends to the south, the transverse polarization at the edges becomes stronger, with higher fractional polarization than that in the jet axis, as shown by the distributions of polarization intensity and fractional polarization along the slice across the local jet (Figure~\ref{fig:AH721}).
This is similar to the sources J0738+1742 \citep{2012evn..confE..94H} and Mrk 501 \citep{2005MNRAS.356..859P}.
The stronger polarization indicates the enhancement of the projected magnetic field components on the plane of the sky \citep{2011MNRAS.415.2081C}.
The perpendicular polarization indicates magnetic field components along the local jet direction dominate beyond the normal components.

Higher fractional polarization at one edge of the jet may result from the interaction with ambient media \citep{2021Galax...9...58G}. 
This would produce a longitudinal B field (i.e. orthogonal linear polarization) \citep{2014MNRAS.437.3405L}.
Possible shocks may arise at the interaction position, compressing the magnetic field at the interaction plane.

Another interesting explanation is the curvature-induced polarization \citep{2013MNRAS.430.1504M,2021Galax...9...58G}.
When the bending happens, the longitudinal magnetic field components ($\langle \textbf{\textit{B}}^{2}_{\rm long}\rangle$) at the jet edge, where the plasma is stretched, increases.
If the local jet is viewed at a large intrinsic view angle, the projected components of $\textbf{\textit{B}}_{\rm long}$ on the plane of the sky would also be enhanced, increasing the fractional polarization.

In this bending region, the position of higher polarization at the north edge shows a significant curved structure. 
However, for the south edge, due to the limited sensitivity, there is only a hint (weak extended emissions) between $1\sim3$ mas that the jet bent up again at the higher polarization position of the south edge (Figure~\ref{fig:AH721}).

To distinguish which origin is more appropriate needs more evidence from multi-frequency observations.
Distributions of spectral index and RM may provide more information.
Both regions indicate that the helical magnetic field may exist with the propagation of the jet at this scale.

\subsection{Magnetic Field at the Hotspots}
Up to $\sim 10$ arcsec scale, the center of the EH shows a stable oblique polarization direction with time (Figures~\ref{fig:AH635} and \ref{fig:AS637 RM}).
Hotspots are formed by the interaction of the jets with the interface with the circumgalactic or intergalactic mediums \citep{2014ARA&A..52..589H}.
The spectral index distribution shows that EH is an optically thin region (Figure~\ref{fig:AS637 RM}).
Thus, the oblique polarization in the center of the EH orients may indicate the oblique interface, where the magnetic field is compressed and ordered at the shock front, contributing to the observed polarization.
As seen in Figure~\ref{fig:AS637 Lband}, the fractional polarization at the edges of EH and WH reach $\sim 30 \%$.

WH has similar polarization distribution and spectral index values (Figrue~\ref{fig:AS637 RM}), leading to the expectation of a symmetrical distribution of the surrounding medium, as the jet is symmetrical at a large scale.
This oblique interface is similar to the expectation for source 3C 273 \citep{2019ApJ...879...75H}.

Transverse RM gradients serve as strong evidence of helical magnetic field even at large scale ( e.g. 5C 4.114 \citep{2015A&A...583A..96G} and Coma A \citep{2017Galax...5...61K}). 
At the EH, we obtained a transverse, sign-change RM gradient (Figrue~\ref{fig:AS637 RM}).
To note, this gradient needs to be confirmed by future observation of upgraded high-sensitivity array VLA \citep{2011ApJ...739L...1P}.

In this region, the termination of the jet, the bulk speed of the outflow decreases to non-relativistic.
The expected polarization for the global helical magnetic field would be toroidal \citep{2002PASJ...54L..39A,2000MNRAS.319.1109G}, contributing to the observed polarization at EH and WH.
The factors affecting the RM values contain the magnetic field and the electron density.
Thus, this gradient may also be the result of the nonuniform distribution of the medium.

Unlike Hydra A, one of the most luminous FR I radio galaxies, which has large RM values ($\sim 10^{3} \rm \ rad \ m^{-2}$) and shows a significant difference in the RM between the approaching lobe and the receding one \citep{1993ApJ...416..554T,2023ApJ...955...16B}, the quasar 1604+159 has similar RM values around the eastern lobe and the western lobe (Figure~\ref{fig:AS637 RM}).
Observed RM values in the approaching lobe and the receding one are affected by viewing angles, environmental differences, or other factors (e.g. the electron density and magnetic field strength in the surrounding medium).
This may suggest that the contribution from the magnetized medium is small and the source is in a different physical condition.

The RM values obtained using the two frequency data have large uncertainties due to the limited samples.
Future wide-band observations are needed to obtain more accurate results. 

\subsection{Emissions from Cocoon ?}

The largest structure detected in this study reaches  $\sim$ 50 arcsec ($\sim$ 300 kilo-parsec), as seen in Figures~\ref{fig:AS637 Lband} and \ref{fig:AS637 RM}.
The source at this scale shows a cocoon-like shape as indicated by the lowest counter line in Figure~\ref{fig:AS637 Lband}.
In addition, several bulges surrounding the core, jets, and hotspots are observed towards the edges, as shown by the green and blue color distributions in the right panel of Figures~\ref{fig:AS637 Lband}.

Polarized emissions are detected in these bulge regions.
After RM correction, the EVPA distribution (Figure~\ref{fig:AS637 RM}) shows similar directions with the uncorrected one (Figure~\ref{fig:AS637 Lband}), which are normal to the edges of the source.
The $L$ band results have higher sensitivity than the RM-corrected because it is the combination of the two frequency data.
Moreover, during the construction of the RM distribution, the two frequency data were clipped to the same UV range to spurious gradients.
Based on our clipping standard (i.e. $I<3\sigma_{I}$, $p<3\sigma_{p}$, and $m<3\sigma_{m}$), the polarization towards the edges (Figure~\ref{fig:AS637 Lband}) is detected at a high significance level.

The surrounding polarized emissions are normal to the edges, indicating the magnetic fields are along the edges.
These emissions have high fractional polarization up to $\sim \ 60 \%$, similar to \citet{1994AJ....108..766B}.
This suggests that the bulges have contributions from non-thermal emission (i.e. synchrotron emission).
Here, we provide a possible explanation for this interesting observation.

In theoretical expectations, there is a simple framework for the way of interaction between the jets from AGNs and the surrounding interstellar medium/intracluster medium (ISM/ICM) \citep{2012rjag.book.....B}.
First, powerful jets resulting from the accretion onto the SMBH propagate with relativistic bulk speed and terminate in regions where hotspots are formed.
Second, backflows are produced by the shocked jets, inflating a cocoon of relativistic plasma with an initial supersonic expanding speed in the ISM/ICM atmosphere.
Third, the interface forms a shell of shocked ISM/ICM, which would be mixed into the cocoon due to the Kelvin-Helmholtz instability.
Fourth, the cocoon expansion stops and finally comes to the `plumes' or `bubbles' of low-density mixed jet/ICM (e.g. Cygnus A\citep{2020ApJ...903...36S} and Hydra A \citep{1993ApJ...416..554T,2023ApJ...955...16B}).

We took the schematic diagram in the 3D relativistic magnetohydrodynamics simulation work \citep{2023MNRAS.526.5418M} to further explain the observed structure.

The electrons, origins of non-thermal radio emission, in different regions obtain energy in their environments due to different acceleration mechanisms \citep{2020NewAR..8901543M}.
At the forward shocks, jets would compress and amplify the external magnetic field in the ambient medium \citep{2023MNRAS.526.5418M}.
The electrons in the magnetized shocked ambient medium (SAM) could produce high synchrotron flux \citep{2023MNRAS.526.5418M}.

The magnetic fields are also amplified and ordered by the shear of the backflows in the cocoon and the compression of the shocks in the contact surface \citep{1990MNRAS.242..616M,2023MNRAS.526.5418M}.
This leads to the magnetic fields along the edges of the cocoon, producing polarization normal to the edges.

The electrons in the cocoon may be more strongly accelerated than in the SAM \citep{2023MNRAS.526.5418M}.
Thus, the synchrotron flux in the cocoon may be stronger than in the SAM.
The observed polarized emissions towards the edges of the large-scale structure are possibly cocoon-dominated.
This could be one possible origin of the strongly polarized emission and high fractional polarization, as observed in Figure~\ref{fig:AS637 Lband}.

The whole cocoon-like shape is similar to source 3C 310 \citep{2012ApJ...749...19K}.
The extended emissions shown by the green and blue colors in the right panel of Figure~\ref{fig:AS637 Lband} may be the ones predicted by \citep{2023MNRAS.526.5418M}, which shows `ring' or `arc-like' high emission structures surrounding the jets towards the edges of the whole cocoon.
These structures possibly arise from the annular shocks in the backflows \citep{2002ApJ...579..176S}.
The 'ring or arc-like' structures are also observed in Hercules A and 3C 310 \citep{2002evn..conf..159G,2002ApJ...579..176S,2012ApJ...749...19K}.

At the largest scale, the eastern part of the source detected more emissions than the western part.
As discussed in \citet{2024ApJ...965...74H}, the viewing angle for the source could be estimated to be $\sim 25^{\circ}$ at parsec scale.
The viewing angle may change to a large value as the jet changes its direction during propagation from parsec to kilo-parsec scale.
In addition, the detection of EH and WH indicates the Doppler effect at the large scale is weaker than at the small scale.
Thus, the whole eastern and western morphology is observed and asymmetrical, while the eastern structure shows more details.

More comprehensive investigations from small-scale to large-scale for the jet-driven AGN feedback, a multi-scale physics, require wideband observations for various sources spanning different stages.
The probe of the observed polarization in the AGN jet increases the wideband configuration requirement for a radio array (e.g. the wideband Jansky VLA \citep{2011ApJ...739L...1P}).
Multi-frequency and multi-epoch observations provide detailed insights into the magnetic field, the spectral, and the ambient medium, leading to more accurate investigations of the origins of the observed emissions.

\section{Conclusions}
We have analyzed the distributions of the total intensity, special index, polarized linear intensity, and Rotation Measure from tens of mas to tens of arcsec scale for the quasar 1604+159.
The source shows remarkable and substantial information on inherent and local magnetic field structures at different scales.
The source was observed at one epoch with the VLBA at the $L$ band and three epochs at the $L$, $X$, and $U$ bands with the VLA.
The conclusions are summarized as follows:
\begin{enumerate}
    \item The jet generally propagates in a collimating structure with several slight bends.
At $\sim$ 10 mas where a standing feature is located, the fractional polarization is $\sim 15\%$.
The jet starts to change its direction from $\sim$ 50 mas.
    \item Perpendicular polarization, which shows the strongest close to the jet axis, was detected from 100 to 500 mas, which may come from the accelerated outflow decreasing the pitch angle of the helical field at the jet rest frame or the velocity shear. 
    \item After that, the jet bends to the south (up to 1 arcsec) with polarization normal to the local jet direction.
As bending, the jet shows strong polarized intensity and high fractional polarization towards the bending edge(s).
Interaction with the ambient medium and/ or curvature-induced polarization are possible origins.
    \item At $\sim$ 20 arcsec scale, the jets propagate along the east-west direction.
Two hotspots are observed at the eastern and western sides of the source, located symmetrically relative to the core.
The flux density ratio ($>1.5$) between the two hotspots suggests the Doppler beaming effect still exists at a large scale.
The polarized emission in the hotspots also shows a symmetrical structure with an oblique direction relative to the jet direction, indicating an oblique magnetic field in the center of the hotspots.
This suggests that the interfaces between the shocks and the ICM may be oblique.
A transverse RM gradient at the eastern hotspot region indicates that a possible helical field at a large scale may also contribute to the observed polarization.
    \item Up to $\sim$ 50 arcsec ($\sim$ 300 kilo-parsec), the source shows a complex structure of the total intensity, with polarization normal to the edges at several bulges and high fractional polarization (up to $\sim 60\%$), indicating highly ordered magnetic fields.
The results indicate that the source may be in the stage when the cocoon expands and shocks, formed during the expansion, heat the ISM/ICM, contributing to the jet-driven AGN feedback.
\end{enumerate}

\section*{Acknowledgements}
\begin{acknowledgments}
We thank the referee for the comments improving the manuscript.
This work was supported by the National Key R$\&$D Programme of China (2018YFA0404602).
L.C. appreciates the support from the National Natural Science Foundation of China (grant 12173066), the National SKA Program of China(Grant No.2022SKA0120102), and the Shanghai Pilot Program for Basic Research – Chinese Academy of Science, Shanghai Branch (JCYJ-SHFY-2021-013).
This research has made use of the NASA/IPAC Extragalactic Database (NED), which is funded by the National Aeronautics and Space Administration and operated by the California Institute of Technology.

\end{acknowledgments}


\vspace{5mm}
\facilities{VLBA, VLA}


\software{astropy \citep{2013A&A...558A..33A,2018AJ....156..123A,2022ApJ...935..167A}
    }




\bibliography{sample631}{}
\bibliographystyle{aasjournal}



\end{document}